\documentstyle[12pt,aaspp4, flushrt]{article}

\def\ls{{_<\atop^{\sim}}}
\def\gs{{_>\atop^{\sim}}}
 
\begin{document}

\title{The variability properties of X-ray steep and X-ray flat
quasars}

\author{Fabrizio Fiore$^{1,2,3}$, Ari Laor$^4$, Martin Elvis$^1$, \\
Fabrizio Nicastro$^{1,2}$, Emanuele Giallongo$^2$}
\affil{$^1$ Harvard-Smithsonian Center for Astrophysics, 
Cambridge MA 02138, USA}
\affil{$^2$ Osservatorio Astronomico di Roma, Monteporzio (Rm) I00040 Italy}
\affil{$^3$ SAX Science Data Center, Roma, Italy}
\affil{$^4$ Physics Department, Technion, Haifa 32000, Israel}
\author{\tt (version: 8pm 11 March 1998) }

\begin{abstract}

We have studied the variability of 6 low redshift, radio quiet `PG'
quasars on three timescales (days, weeks, and months) using the ROSAT
HRI.  The quasars were chosen to lie at the two extreme ends of the
ROSAT PSPC spectral index distribution and hence of the H$\beta$ FWHM
distribution.  The observation strategy has been carefully designed to
provide even sampling on these three basic timescales and to provide a
uniform sampling among the quasars

We have found clear evidence that the X-ray steep, narrow H$\beta$,
quasars systematically show larger amplitude variations than the X-ray
flat broad H$\beta$ quasars on timescales from 2 days to 20 days. On
longer timescales we do not find significant differences between steep
and flat quasars, although the statistics are poorer.  We suggest that
the above correlation between variability properties and spectral
steepness can be explained in a scenario in which the X-ray steep,
narrow line objects are in a higher $L/L_{\rm Edd}$ state with respect
to the X-ray flat, broad line objects.
 
We evaluated the power spectrum of PG1440+356 (the brigthest quasar in
our sample) between $2\times10^{-7}$ and $\sim10^{-3}$ Hz, where it
goes into the noise.  The power spectrum is roughly consistent with a
1/f law between $10^{-3}$ and $2\times10^{-6}$ Hz.  Below this
frequency it flattens significantly.

\end{abstract}


\section{Introduction}
\bigskip

While X-ray variability in Seyfert galaxies has been the subject of
intensive study (see review by Mushotzky, Done \& Pounds 1993, and
Green et al 1993, Nandra et al. 1997), the variability of quasars,
being fainter, have received far less attention (the Zamorani et al.
1984 study remains the most extensive to date).  Higher luminosity AGN
had been expected to be physically larger and so have slower, and
most likely lower amplitude, variations. A typical quasar is 100--1000
times more luminous than the highly variable Seyferts, such as
NGC~4051, and so would vary at a significantly lower rate, i.e. a
minimum of several days. It was probably this expectation that
deterred extensive observing campaigns.  This is unfortunate since we
show here that X-ray variability is common, rapid and of quite large 
amplitude ($\gs$ factor of 2). The variability most
likely originates in the innermost regions of quasar, and so can help
unravel the basic parameters of the quasar central engine (mass,
geometry, radiation mechanisms, radiative transfer) none of which are
yet well constrained.

These speculations are supported by compact galactic sources, for
which the investigation of the ``variability properties vs. spectral
shape'' and the ``variability properties vs. luminosity'' planes have
brought a great improvement in our understanding (see e.g. van der
Klis 1995). The same is likely to happen for quasars.  Compact
galactic sources are usually bright and variable on timescales
as short as 1 ms,
which means that their variability timescales can be probed by just a
few observations of individual objects.  For example, the Galactic
black hole candidate (BHC) Cyg X-1 recently underwent two dramatic
changes in its variability and spectral properties (see e.g. Cui et
al. 1997), and it was possible to follow the whole cycle from the
usual low and hard state, to a medium-high, softer state, and back to
the low and hard state in the course of a few months.  This is
unlikely to happen in a luminous AGN, even if there were similarities
between AGN and black hole candidates, because the variability
timescales (and the sizes and luminosities) are probably much longer
(greater and higher) in AGN.  Instead, if the analogy between quasars and
BHC holds, we will observe a population of quasars some in a `high and soft'
state, and some in a `low and hard' state.

Hence the analogous observational way forward in quasars is to investigate the
variability properties of samples of quasars, selected
according to their spectral properties and luminosity.

The evidence for rapid, large amplitude, X-ray variability
in a few AGN with unusually steep X-ray spectra and unusually narrow
Balmer lines (mostly Narrow Line Seyfert 1 galaxies, NLSy1, 
given the strong correlation between these two
quantities found by Laor et al., 1994, 1997, Boller, Brandt \& Fink 1996)
has recently grown:

\begin {enumerate}
\item
NGC4051 shows large variations (50 \%) on timescales of $\approx 100$
seconds and has very narrow optical and UV emission lines and steep
0.1-2 keV X-ray spectrum. It is also highly variable in the EUV 
(a factor of factor of 20 in 8 hours, Fruscione et al 1998).

\item
IRAS13224-3809 has a
particularly steep 0.1-2 keV spectrum ($\alpha_X>3$)
narrow H$\beta$ line, very strong FeII emission and shows X-ray
variations of a factor 50 or more on timescales of a few days
(Otani 1995, Brandt et al. 1995, Boller et al 1997) and 
a factor of 2 in less than 800 seconds (Boller et al. 1996);

\item
RE J 1237+264 showed one very large (factor of 50) variability event
(Brandt, Pounds \& Fink, 1995)

\item
PHL1092 varied by a factor of 4 in 2 days (Forster \& Halpern 1996, 
Lawrence et al. 1997). PHL1092 is a 
high luminosity ($5\times10^{46}$ erg s$^{-1}$) very steep
0.1-2 keV spectrum and narrow emission line quasar.

\item
The relatively high luminosity quasar NAB0205+024 
($L_{0.5-10 keV}=8\times10^{44}$ erg s$^{-1}$) 
shows variations of a factor of 2 in less than 20 ks in an ASCA
observation (Fiore et al. 1998a). 

\item
Mark 478 (PG1440+356) shows factor of 7 variations in 1 day during a 
long EUVE monitoring (Marshall et al 1996)

\item
The extremely soft NLSy1 WPVS007 ($\alpha_{0.1-2keV}=7.3$)
showed a huge variation (factor of 400) 
variation between the RASS observation and a follow-up PSPC
pointed observation taken 2 years later (Groupe et al. 1995).


\end{enumerate}

At this point we believe that a systematic study of a sample of normal
quasars spaning the observed range of properties is needed.  A
systematic study of the variability properties of AGNs is not an easy
task, since it requires:

\begin {enumerate}
\item
the selection of well defined and representative (i.e. unbiased) sample;

\item
the availability of numerous repeated observations;

\item
a carefully designed observational strategy. In particular, it is very
important that the sampling time is regular and that it is similar for
all objects in the sample, so that the results on each object can be
easily compared with each other. Otherwise the differences in the
observed variability may be induced just by differences in sampling
patterns.

\end{enumerate}

Because of these stringent requirements systematic studies are still lacking
(for example the Zamorani et al. 1984 study on quasars, and the Green
et al. 1993 study on Seyfert galaxies do not fulfill any of the three
criteria above).  To explore in a systematic manner the possible relation
between line width, soft X-ray spectrum and X-ray variability
properties, we initiated a pilot program using the ROSAT HRI.  In the
following we present the results from a campaign of observations of
six PG quasars which addresses the above points.

\section{The sample}

Six quasars were chosen from a complete sample of 23 optically
selected (PG) quasars ($M_B<-23$), all studied with the ROSAT PSPC
(Laor et al. 1994, 1997).  To maximize the rest frame soft X-ray
detectability in the whole 0.1-2 keV PSPC band the Laor et al. sample
of PG quasars ($M_B<-23$) is limited to quasars with: low redshift
(z$<$0.4) and low Galactic N$_H$ ($< 1.9\times 10^{20}$cm$^{-2}$).
This sample of AGN has the advantage that it is extracted from a
complete sample of optically selected quasars, and should therefore be
representative of such quasars.  In particular, the Laor et al. sample
selection criteria are independent of the X-ray properties, and thus
the sample is free from X-ray selection biases.

\begin{figure}
\epsfysize=15cm 
\epsfbox{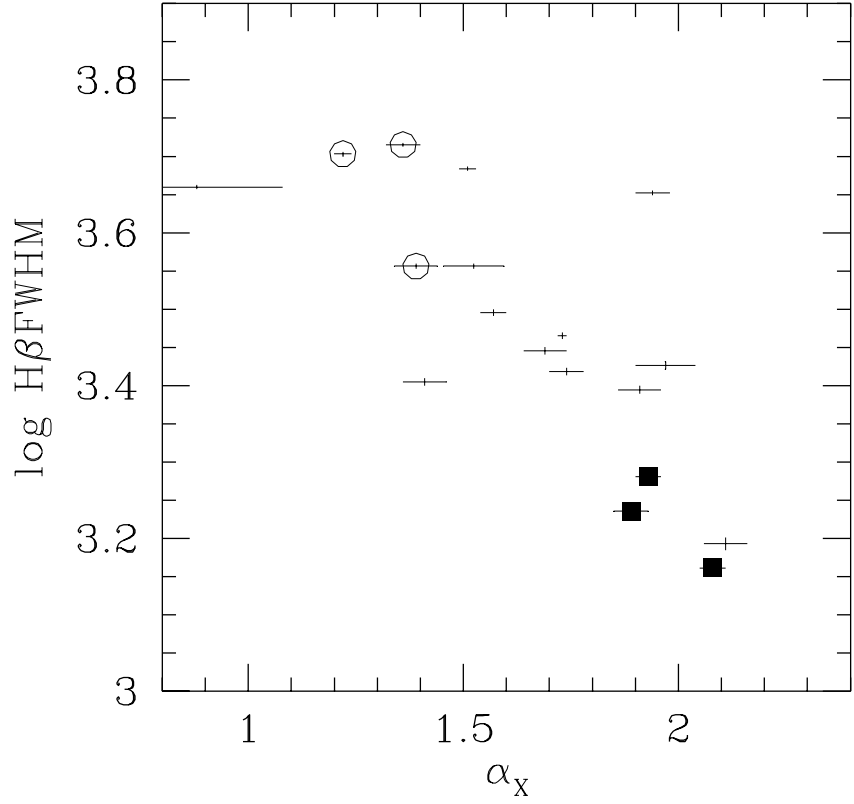} 
\figcaption{
\label{alpha-hbeta}
H$\beta$ FWHM as a function of 
$\alpha_X$ for the quasars in the Laor et al sample. 
Filled squares and open circles identify the three X-ray steep and the
three X-ray flat quasars discussed in this paper. 
}
\end{figure}

The six quasars were chosen to lie at the two extreme ends of the PSPC
spectral index distribution ($f_{\nu}\propto \nu^{-\alpha_X}$, 
$1.2\ls \alpha_X \gs2.1$)
and hence of the H$\beta$ FWHM distribution.  Figure
\ref{alpha-hbeta} show the H$\beta$ FWHM as a function of $\alpha_X$
for the radio quiet quasars in the Laor et al sample.  (We excluded
the X-ray weak quasar PG1001+054 which is too faint for followup
variability studies with the HRI).  X-ray steep and flat quasars
discussed in this paper are identified by filled squares and open
circles respectively.  The quasar with $\alpha_X<1$ and large errors
is PG1114+445, which has an ionized absorber along the line of sight
(Laor et al. 1997, George et al. 1997).

Objects with extreme properties should help highlight the
basic correlations which may provide important hints for the
underlying physical processes. Table 1 gives the
redshift, the PSPC energy index, the 2 keV and 3000 \AA\ luminosities
(for $H_0=50$~km~s$^{-1}$~Mpc$^{-1}$, $q_0=0.5$)
and the $H_{\beta}$ FWHM for the six quasars.
All the chosen quasars are radio-quiet.  

\begin{table}[h]
\caption{Continuum and Line Parameters}
\begin{tabular}{lccccc}
\hline
name & z & $\alpha_X$ & $\nu L_{\nu}(2keV)$ & $\nu L_{\nu}(3000\AA)$
& H$\beta$FWHM \\
     &   &   & erg s$^{-1}$        & erg s$^{-1}$  &  km s$^{-1}$  \\
\hline
\multicolumn{6}{c}{Steep X-ray quasars} \\
PG1115+407 & 0.154 & 1.9      & 43.61  & 44.96 & 1720  \\
PG1402+261 & 0.164 & 1.9      & 44.08  & 45.45 & 1910   \\
PG1440+356 & 0.077 & 2.1      & 43.64  & 44.91 & 1450  \\
\hline   					       
\multicolumn{6}{c}{Flat X-ray quasars} \\		       
PG1048+342 & 0.167 & 1.4      & 43.85  & 45.14 & 3600  \\
PG1202+281 & 0.165 & 1.2      & 44.25  & 45.00 & 5050  \\
PG1216+069 & 0.334 & 1.4      & 44.78  & 45.94 & 5190  \\
\hline
\end{tabular}
\end{table}

\section{The observational strategy}
 
The observational strategy was designed to: a) provide even sampling
on three basic timescales (days, weeks, and months), b) provide a
uniform sampling among the quasars, and c) keep the total integration
time reasonably small.  These constraints result in a
quasi-logarithmic observation plan: one $\sim2000$ s exposure a day
for one week; one $\sim 2000$ s exposure each week for one month; two
$\sim2000$ s exposures a month apart, 6 months after the main
campaign, i.e.  6-7 points with an {\it even} spacing of 1 day; 4-5
points with an {\it even} spacing of 1 week 2 points with a spacing of
1 month and 2 sets of 2 points with a spacing of 6 months.  The S/N
obtained allows the $3\sigma$ detection of $\sim 30 \%$ variability in
the faintest source of the sample.

\section{Results}

Tables 2 and 3, respectively, give the observation date, the net HRI
exposure and the background subtracted HRI count rate for the X--ray
steep and flat quasars.  Counts were extracted from a region of 60
arcsec radius. Since $\sim 90 \%$ of the counts from a point source
fall within the same extraction region (David et al. 1997) no
correction has been made for lost flux.  Background was extracted from
two 60 arcsec radius regions straddling the source region. The
background is small in all cases ($< 10 $ counts ks$^{-1}$ in a 60
arcsec radius regions).

\begin{table}[h]
\caption{HRI Monitoring of 3 X-ray Steep Quasars}
\begin{tabular}{lccccc}
\hline
Source & Observation Date & Exposure$^a$ & HRI Count-rate$^b$ \\
\hline
& 07/Nov/95 & 3.14 & $87\pm6$ \\
& 09/Nov/95 & 2.39 & $80\pm6$ \\
& 10/Nov/95 & 2.23 & $83\pm7$ \\
& 11/Nov/95 & 2.29 & $87\pm7$ \\
PG 1115+407 &  12/Nov/95 & 2.91 & $101\pm6$ \\
& 13/Nov/95 & 3.53 & $106\pm6$ \\
& 14/Nov/95 & 3.62 & $70\pm5$ \\
& 21/Nov/95 & 1.71 & $65\pm7$ \\
& 29/Nov/95 & 2.06 & $53\pm6$ \\
& 05/May/96 & 2.00 & $139\pm9$ \\
\hline
& 05/Jan/96 & 1.10 & $380\pm20$ \\
& 11/Jan/96 & 4.40 & $329\pm9$ \\
&  12/Jan/96 & 3.30 & $422\pm12$ \\
PG 1402+261 & 14/Jan/96 & 4.33 & $331\pm9$ \\
& 15/Jan/96 & 3.24 & $304\pm10$ \\
& 18/Jan/96 & 3.84 & $170\pm7$ \\
& 25/Jan/96 & 4.80 & $204\pm7$ \\
& 20-21/Jun/96 & 3.86 & $201\pm8$ \\
\hline
& 03/Jan/96 & 3.69 & $1339\pm19$ \\
& 04/Jan/96 & 0.56 & $770\pm40$ \\
& 05/Jan/96 & 2.66 & $847\pm18$ \\
& 06/Jan/96 & 4.02 & $880\pm15$ \\
PG 1440+356 & 07/Jan/96 & 4.35 & $900\pm15$ \\
& 08/Jan/96 & 3.99 & $596\pm13$ \\
& 10/Jan/96 & 5.51 & $1304\pm16$ \\
& 16/Jan/96 & 3.51 & $800\pm15$ \\
& 23/Jan/96 & 3.50 & $561\pm13$ \\
& 20/Jun/96 & 2.72 & $1276\pm22$ \\
\hline
\end{tabular}

$^a$ in ks; $^b$ in ks$^{-1}$
\end{table}

\begin{table}[h]
\caption{HRI Monitoring of 3 X-ray Flat Quasars}
\begin{tabular}{lccc}
\hline
Source & Date Obs & Exposure$^a$ &HRI Count-rate$^b$ \\
\hline
& 07/Nov/95 & 4.13 & $105\pm5$ \\
& 08/Nov/95 & 2.08 & $115\pm8$ \\
& 09/Nov/95 & 2.06 & $125\pm8$ \\
& 10/Nov/95 & 2.14 & $106\pm8$ \\
PG 1048+342 & 11/Nov/95 & 3.30 & $103\pm6$ \\
& 14/Nov/95 & 3.40 & $78\pm5$ \\
& 21/Nov/95 & 1.86 & $91\pm8$ \\
& 28-29/Nov/95 & 3.02 & $103\pm6$ \\
& 04/May/96 & 4.31 & $87\pm5$ \\
& 02/Jun/96 & 2.54 & $43\pm5$ \\
\hline
& 25/Nov/95 & 2.59 & $153\pm8$ \\
& 02/Dec/95 & 3.51 & $173\pm8$ \\
& 03/Dec/95 & 2.19 & $167\pm10$ \\
& 03-04/Dec/95 & 2.37 & $164\pm9$ \\
PG 1202+281 & 04-05/Dec/95 & 2.32 & $151\pm9$ \\
& 06/Dec/95 & 2.09 & $149\pm9$ \\
& 07/Dec/95 & 2.33 & $144\pm9$ \\
& 08/Dec/95 & 2.84 & $156\pm8$ \\
& 09/Dec/95 & 3.03 & $140\pm7$ \\
& 16/Dec/95 & 2.10 & $197\pm11$ \\
& 23/May/96 & 3.50 & $99\pm6$ \\
& 21/Jun/96 & 4.56 & $62\pm4$ \\
\hline
& 09/Dec/95 & 2.46 & $67\pm6$ \\
& 10/Dec/95 & 2.33 & $74\pm6$ \\
PG 1216+069 & 11/Dec/95 & 2.65 & $76\pm6$ \\
& 12/Dec/95 & 3.16 & $67\pm5$ \\
& 17/Dec/95 & 3.80 & $69\pm5$ \\
& 23/Dec/95 & 0.88 & $67\pm10$ \\
& 06/Jul/96 & 3.79 & $80\pm5$ \\
\hline
\end{tabular}

$^a$ in ks; $^b$ in ks$^{-1}$
\end{table}

We have also looked at longer variability timescales (of the order of
2000 days) by comparing the HRI fluxes with the PSPC fluxes obtained
during the Rosat All Sky Survey (RASS) and the pointed ROSAT PSPC
observations reported in Laor et. (1997).  We converted the PSPC count
rates to effective HRI count rates by assuming the spectral shape
observed in the ROSAT pointed observations (Laor et al. 1997).  RASS
and pointed PSPC observations `converted' count rates are given in
Table 4.

\begin{table}[h]
\caption{RASS and PSPC pointed observation}
\begin{tabular}{lcc}
\hline
Source & RASS & PSPC pointed \\
Count-rate$^a$ & Date Obs & Count-rate$^a$ \\
\hline
PG1115+407 & 108$\pm$9 & 65$\pm$5 \\
PG1402+261 & 133$\pm$10 & 195$\pm$4 \\
PG1440+356 & 1028$\pm$18 & 367$\pm$6 \\ 
\hline   
PG1048+342 & 56$\pm$6 & 51$\pm$3 \\
PG1202+281 & 152$\pm$10 & 107$\pm$3 \\
PG1216+069 & 96$\pm$8 & 99$\pm$4 \\
\hline
\end{tabular}

$^a$ `converted' (see text) PSPC count rates in ks$^{-1}$ 

\end{table}

\subsection {Logarithmic normalized light curves}

\begin{figure}
\epsfysize=15cm 
\epsfbox{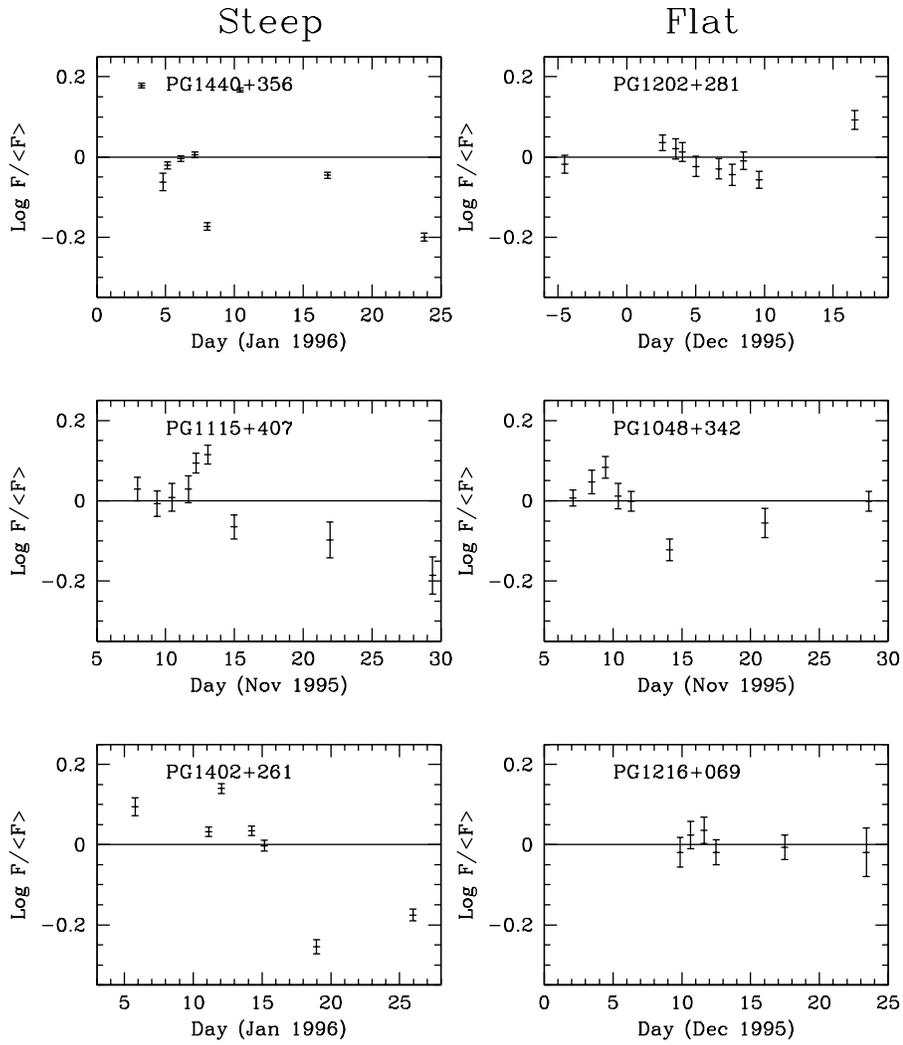} 
\figcaption{
\label{lcshort}
The ROSAT HRI light curves (plotted as logarithm of
the ratio of the flux to the average flux) of the six quasars on a 2-20
day timescale.
}
\end{figure}

\begin{figure}
\epsfysize=15cm 
\epsfbox{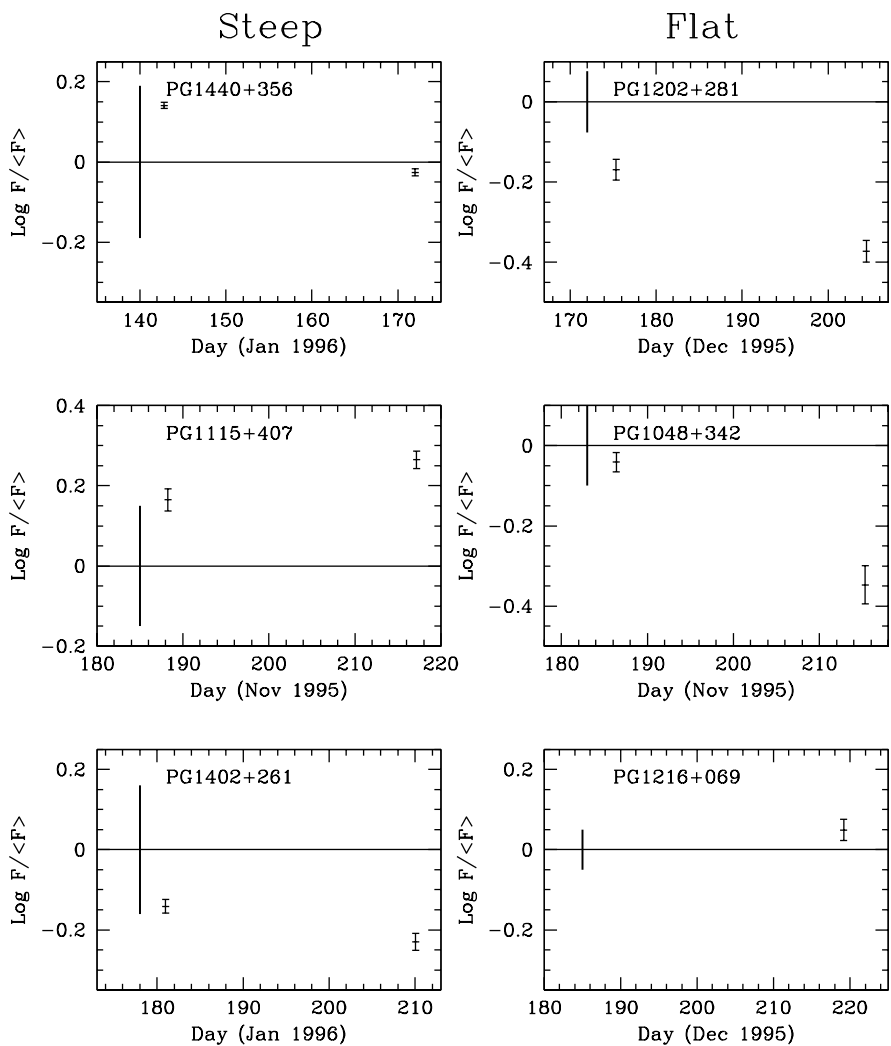} 
\figcaption{\label{lclong}
The HRI observations of the six quasars 
performed six months later compared with the average flux and 
the dispersion observed on timescales shorter than one month
(20-200 days timescale).}
\end{figure}

\begin{figure}
\epsfysize=15cm 
\epsfbox{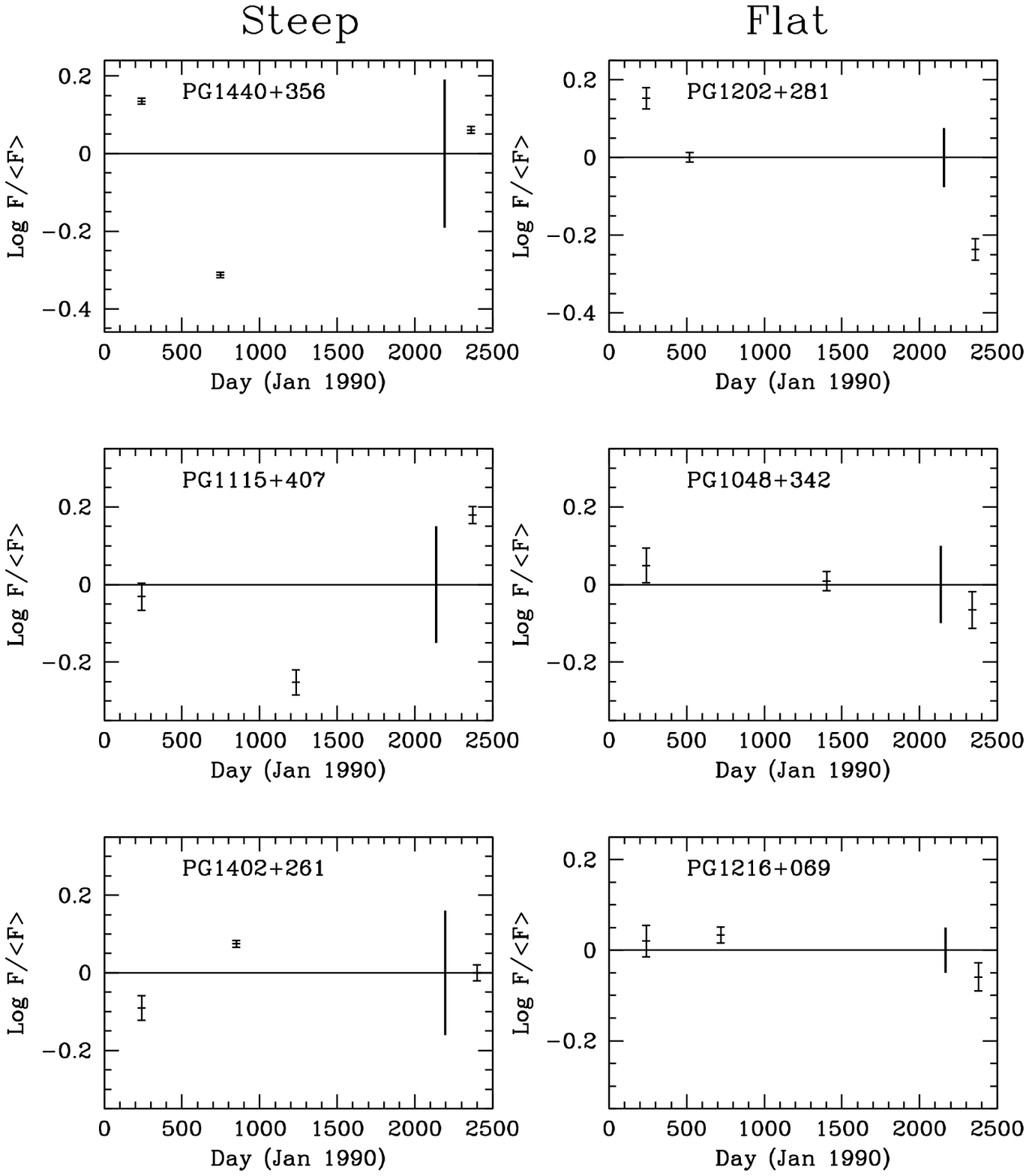} 
\figcaption{\label{lcvlong}
The HRI observations of the six quasars 
compared with the RASS and with the PSPC pointed observations
(200-2000 days timescale).}
\end{figure}

Figure \ref{lcshort} shows the logarithm of the light curves of the
six quasars, each normalized to the mean flux, spanning about one
month each, while figure \ref{lclong} shows the observations performed
six months later compared with the average flux and the dispersion
observed on timescales shorther than one month.  Figure \ref{lcvlong}
shows the normalized RASS and PSPC pointed observation points, the
dispersion as in Figure \ref{lcshort} and the normalized HRI
observation performed 6 months later.

The steep $\alpha_X$ quasars show large variations (up to a factor of
2) on all the timescales investigated. Conversely, the flat $\alpha_X$
quasars show little variability on short timescales. Factor of 2
variations with respect to the average flux seem to be allowed for
both X--ray steep and X--ray flat quasars on 200-2000 days timescales,
although the statistics on these timescales are poorer.

\subsection {Structure function}

To compare the variability of the quasar groups more quantitatively we
computed the logarithm of the ratio between each pair of flux
measurements (at two times $t_i$ and $t_j$ in the quasar rest
frame). We then computed the median and the mean in 9 time bins for
both flat and steep X-ray quasar samples:

$$<\Delta m>= median~or~mean ( |2.5*log(f(t_j)/f(t(i)))|). \leqno(1) $$

\noindent

The function using the mean is similar to the so 
called `average structure function' (see e.g. Di Clemente et al. 1996).

We have verified that no single quasar dominates the median and mean 
in eq. (1).

\begin{figure}
\epsfysize=15cm 
\epsfbox{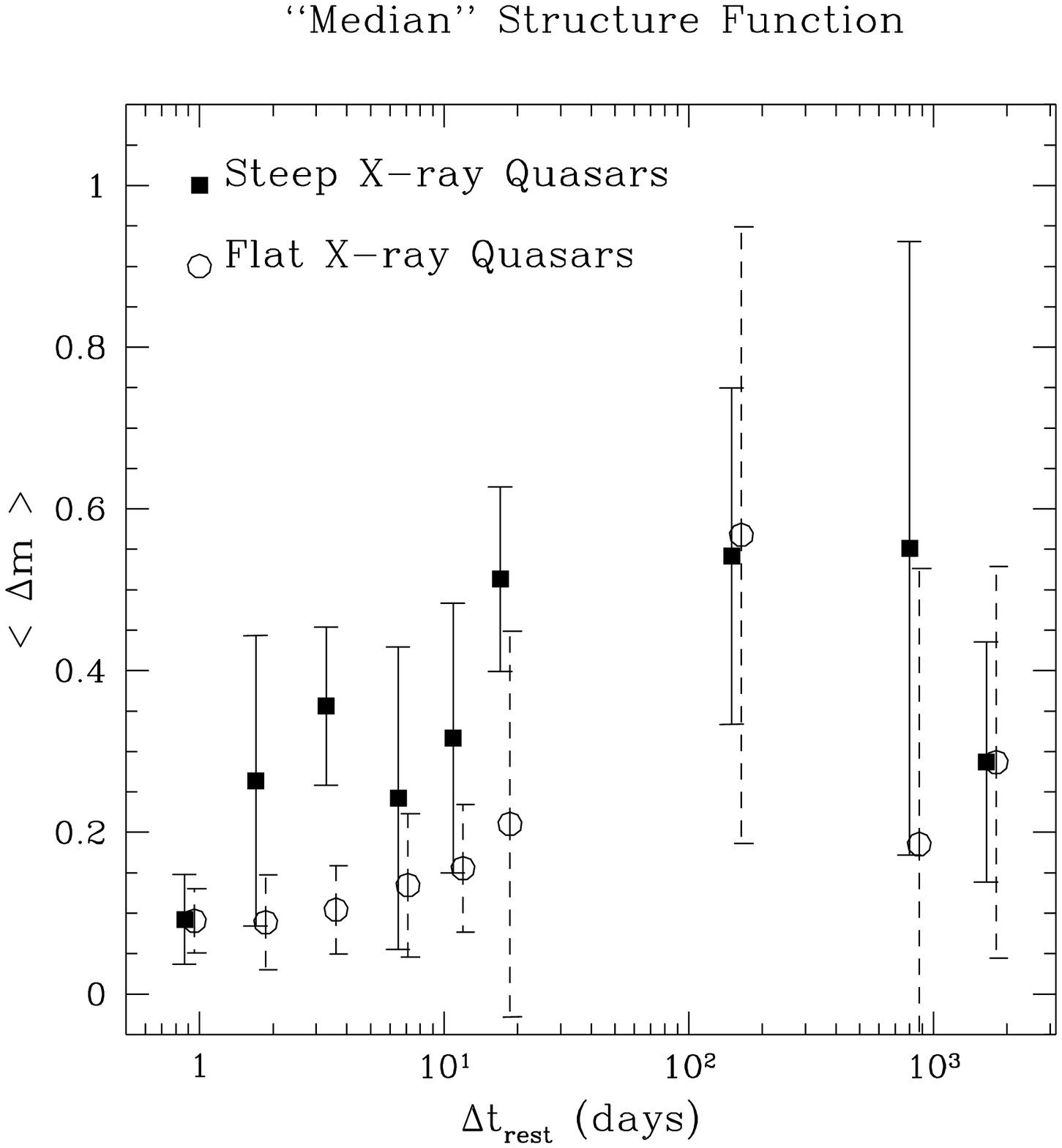} 
\figcaption{\label{stru_med}
``Median'' Structure Function ($<\Delta m>$, see eq. 1) for the X-ray steep 
(filled squares) and flat (open circles). X-ray flat quasars points have been 
slightly shifted from real $\Delta t$ for a sake of clarity.
}
\end{figure}

\begin{figure}
\epsfysize=15cm 
\epsfbox{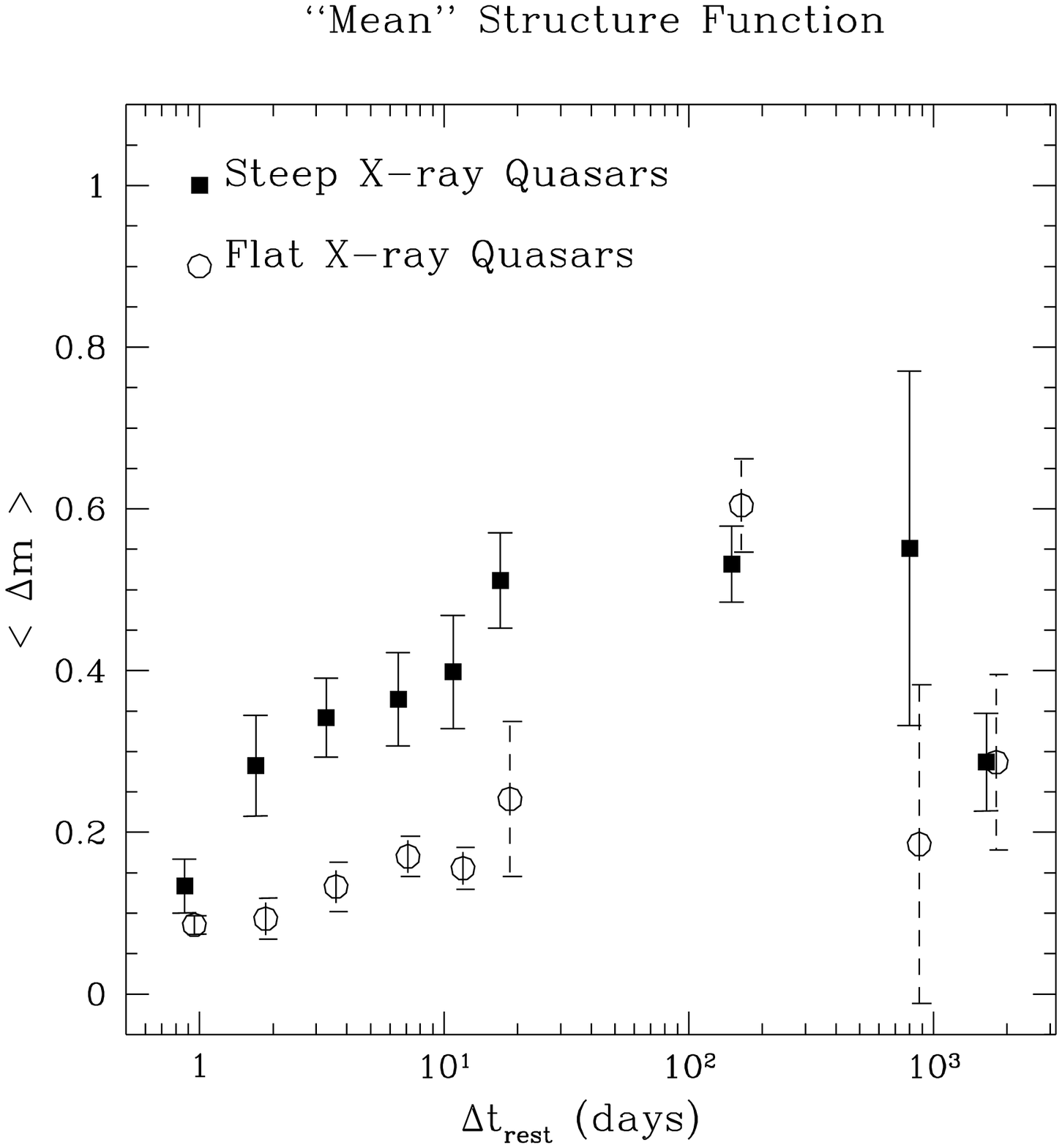} 
\figcaption{\label{stru_mean}
Mean Structure Function ($<\Delta m>$, see eq. 1) 
for the X-ray steep (filled squares) and flat (open circles). 
X-ray flat quasars points have been 
slightly shifted from real $\Delta t$ for a sake of clarity.
}
\end{figure}

$<\Delta m>$ for the median is shown in figure \ref{stru_med}.
Error bars represent here the semi interquartile range.
$<\Delta m>$ for the mean is shown in figure \ref{stru_mean}.
Error bars represent here the error on the mean and are
therefore less conservative.

It is clear that the $<\Delta m>$ (both mean and median) for X-ray
steep quasars is significantly higher than that of X-ray flat quasars
on timescales from 2 days to 20 days.  On the longest timescale (200
days) the two $<\Delta m>$ are consistent with each other, with large
errorbars.  The point at about 2000 days in figures \ref{stru_med} and
\ref{stru_mean} compares the HRI monitoring results with the RASS
results.  The point at about 1000 days compares the HRI monitoring and
the RASS results with the ROSAT PSPC pointed observations. The error
bars here are larger because of the different sampling times used.

\subsection {Power spectrum of PG1440+356}

\begin{figure}
\epsfysize=15cm 
\epsfbox{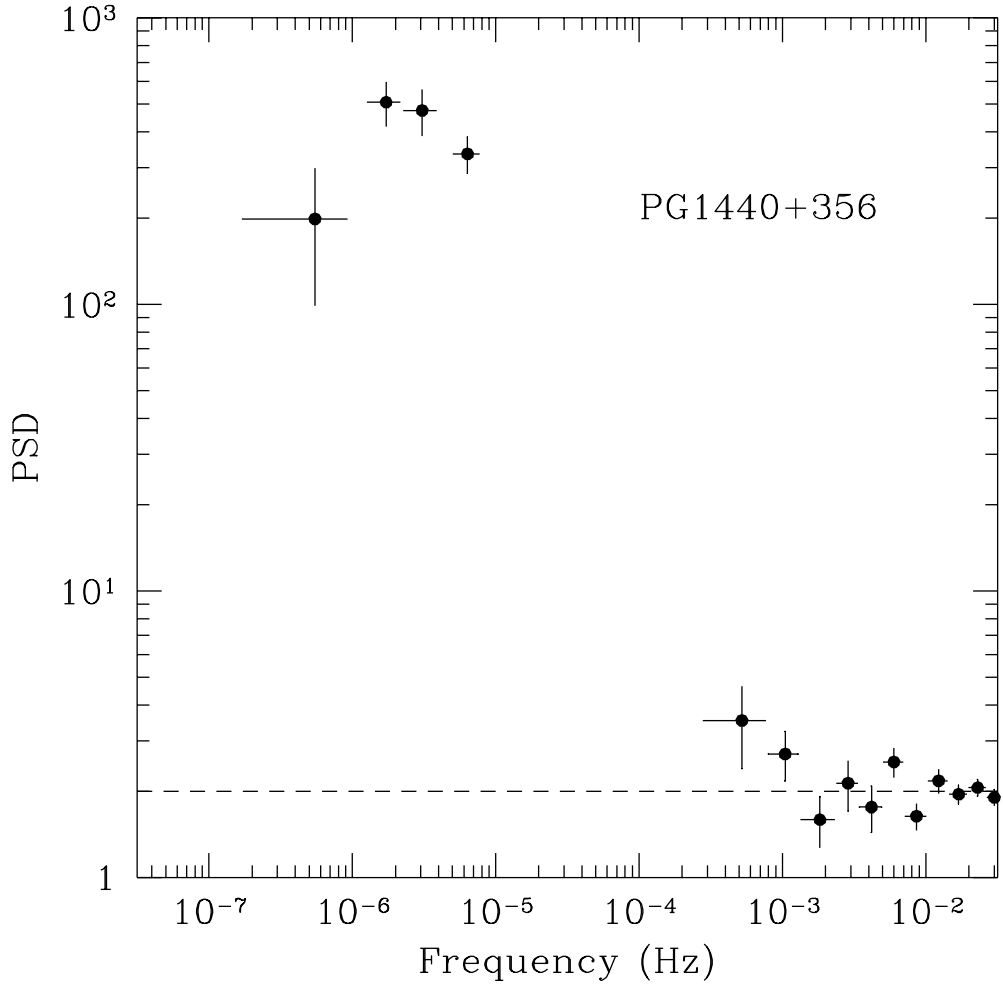} 
\figcaption{\label{pds}
The power spectrum of PG1440+356 from 0.03 to $2\times10^{-7}$ Hz.
The power spectrum is normalized such that the level of the noise 
is 2 (dashed line).
}
\end{figure}

For the brightest quasar in our sample (PG1440+356) it is possible to
investigate timescales shorter than the duration of each single HRI
exposure.  Analysis of the HRI light curves of each single OBI reveals
significant variability down to a time scale of a few hundred
seconds. To quantify the amplitude of this variability we have
computed a power spectrum for each of the single HRI exposures and, to
improve the statistics, we have added together the 12 power spectra.
The resulting coadded power spectrum is shown in figure \ref{pds},
along with the lower frequency power spectrum calculated at four
reasonably well sampled frequencies using the full HRI monitoring.
The error bars represent the dispersion of the power in each bin.

The power spectrum above $10^{-3}$ Hz is flat, and thus consistent
with white noise. Excess power is detected at $\sim5 \times10^{-4}$
Hz, which is a factor $\sim 100$ smaller than that detected at $\sim5
\times10^{-6}$ Hz, a mean 1/f distribution. Unfortunately we do not
have information on the power in the decade between $10^{-5}$ and
$10^{-4}$ Hz and therefore we cannot tell whether the power spectrum
decreases smoothly as 1/f above $\sim5 \times10^{-6}$ Hz or rather if
it breaks at an intermediate frequency.  Continuous monitoring up to
several days is needed to answer to this question.  

The uncertainty on the power below $10^{-6}$ Hz is large and then we
cannot assess whether the power spectrum stays constant below this
frequency or rather decreases, as suggested by the low frequency point
in figure \ref{pds}. We can however exclude that the power spectrum
continues to increase as 1/f below $10^{-6}$ Hz, although the precise
break frequency is ill defined.  Systematic monitoring on weeks to
months timescales can strongly constrain this frequency.  In the
framework of accretion disks models, this timescale may be related
with the thermal or viscous instabilities, and its measure could help
to estimate important model parameters like the mass of the central
black hole and the viscosity parameter $\alpha$ (e.g. Siemiginowska \&
Czerny 1989)

\section{Comparison with other work}

Nandra et al. (1997) used the source root mean square variation
$\sigma^2_{rms}$ from ASCA SIS data for a sample of (X-ray bright) AGN
to suggest an inverse relation between variability and X-ray
luminosity. Our sample of high luminosity, optically selected objects
allows us to test this suggestion.  We have calculated the noise
subtracted $\sigma^2_{rms}$ according to the method of Nandra et al
(1997).  These are reported in Table 5 for two different time scales:
days and months-to-years.  (We do not show errors on $\sigma^2_{rms}$
because the analytic formula provided by Nandra et al. 1997 is not
valid in the limit of small number of points.)  The light curve of
PG1216+069 in \ref{lcshort} is consistent with a constant value and
therefore $\sigma^2_{rms}$ is formally zero in this case. We take the
value of $\sigma^2_{rms}$ calculated for the other flat X-ray quasars
($\approx 0.01$) as an upper limit of $\sigma^2_{rms}$ in PG1216+069.

\begin{table}[h]
\caption{Source root mean square variation}
\begin{tabular}{lcc}
\hline
name  & $\sigma^2_{rms}$(2-20 days) & $\sigma^2_{rms}$(200-2000 days) \\
      &  $10^{-2}$ & $10^{-2}$ \\
\hline
\multicolumn{3}{c}{Steep X-ray quasars}\\
PG1115+407 &  3.2 & 13.8 \\ 
PG1402+261 &  7.3 & 19.1 \\ 
PG1440+356 &  8.3 & 15.2 \\ 
\hline                        
\multicolumn{3}{c}{Flat X-Ray quasars}\\
PG1048+342 &  1.2 & 13.9 \\ 
PG1202+281 &  0.7 & 9.7  \\ 
PG1216+069 &  0.0 & 1.8 \\  
\hline
\end{tabular}
\end{table}

\begin{figure}
\epsfysize=15cm 
\epsfbox{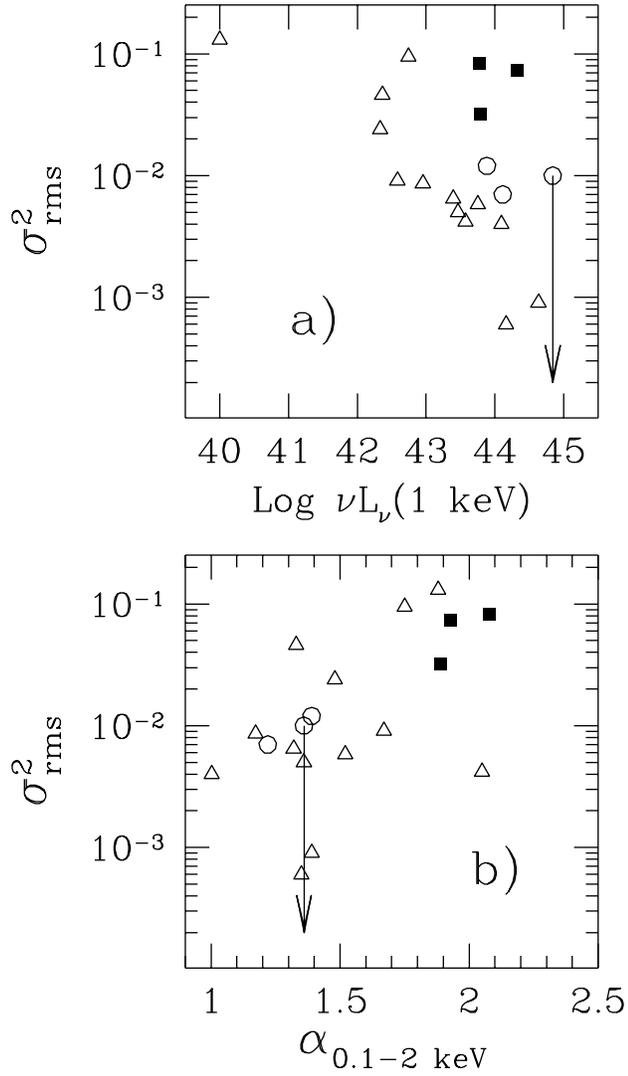} 
\figcaption{
\label{rms}
The `excess' variance as a function of the 1 keV luminosity 
(in erg s$^{-1}$ Hz$^{-1}$) and of the 0.1-2 PSPC spectral index.
Open triangles show quasars and Seyfert galaxies from Nandra et al. 1997,
filled squares identify steep X-ray spectrum quasars, 
open circles identify  flat X-ray spectrum quasars.
}
\end{figure}

Figure \ref{rms} shows $\sigma^2_{rms}$ as a function of the X-ray
luminosity and of the soft (0.1-2 keV) X-ray spectral index for the
six quasars in our study and for the sources in Nandra et al (1997)
with no strong low energy absorption (to provide a low energy X-ray
spectral index). For this figure we used the $\sigma^2_{rms}$ SIS
0.5-2 keV in Nandra et al (1997). This band overlaps quite well with
the HRI band (the HRI has some response in the `carbon' band, below
0.3 keV, but this is much smaller than in the 0.5-2 keV band).  The
PSPC spectral indices for the Nandra et al.  objects are taken from
Walter \& Fink (1993) and Laor et al (1997) or are calculated from
WGACAT hardness ratios when not available in these papers. The
variability of the Nandra et al. objects was measured on a time scale
of $\sim 1-4\times 10^4$~s, i.e. about an order of magnitude shorter
than the timescales probed for the PG quasars (for the six PG quasars
$\sigma^2_{rms}$ in figure\ref{rms} refers to the 2-20 days time
scale).

The addition of the six quasars makes a substantial difference. For the 
Nandra et al. sample alone the $\sigma^2_{rms}$ vs $L_{1keV}$ linear
correlation coefficient is r=--0.82 (probability of 99.9 \%), while
with the quasars it is reduced to r=--0.54. This is still significant
(probability of 95 \%),
although it depends entirely on one point (the low luminosity Seyfert
1 galaxy NGC4051). The values of $\sigma^2_{rms}$ and $\alpha_X$ were
uncorrelated for the Nandra et al. sample alone 
(r=+0.37, probability of 15\%); adding the quasars results in r=+0.55, a
95 \% probability correlation.



At a given luminosity the three x-ray `steep' quasars from our sample
all have $\sigma^2_{rms}$ higher than both the `flat' X-ray quasars in
our sample and all the other objects in the Nandra et al.  (1997)
sample by a factor of 10 to 100.  The $\sigma^2_{rms}$ for the
Nandra et al. (1997) sample refers to a factor of ten smaller
timescale, and this might explain why they show lower amplitude
variability for the same X-ray luminosity. However, 
if the power spectrum of the steep X-ray spectrum
quasars falls like $\approx 1/f$ on the 0.1 to 1 day timescales
(as suggested for PG1440+356, see \S
4.3), then only part of the difference in $\sigma^2_{rms}$ could be
explained. If the power spectrum stays more or less constant on
these timescales then this difference is highly significant.  A
comparison of variability amplitude in this frequency range between
broad and narrow line Seyfert galaxies using ASCA and BeppoSAX light
curves of similar sampling and length is in progress and will be part
of a forthcoming paper (Fiore et al. 1998, in preparation).

Finally we note that the 1 keV X-ray luminosity and the spectral index
are slightly anti-correlated (r=--0.20) in the sense that steeper
objects tend to have lower 1 keV luminosities. The correlation is not
significant here, but Laor et al. (1997) find a more significant
correlation between $\alpha_X$ and the 2 keV luminosity in their
complete, optically selected 
sample of PG quasars ($r_s=0.482$, probability of 98 \%). Thus, low
luminosity AGNs tend to be more variable because of the
luminosity-variability correlation, but on the other hand
they also tend to have flatter
spectra, which diminishes their variability due to the spectral
index-variability correlation. To remove the effect of the
$\alpha_X-L_x$ correlation from the
$\sigma^2_{rms}-L_x$ and $\sigma^2_{rms}$--$\alpha_X$
correlations we performed a partial correlation analysis (e.g.,
Kendall \& Stuart 1979).  The correlation coefficients turns out to be
very similar to the previous ones (-0.53 and 0.54), giving a
probability for a correlation again of 95 \%.  This is still a
marginal result, which could be highly strengthened or weakened by
adding just a few more points at low and high luminosities and at flat
and steep $\alpha_X$.

A correlation between variability and spectral shape, similar to that
discussed in this paper, has been reported also by Green et al. (1993)
and K\"onig, Staubert and Timmer (1997) in two different analyses of
EXOSAT lightcurves of AGN.

Di Clemente et al. (1996) studied the variability of 30 PG quasars in
the red band and compared it with that of a sample of 21 PG quasars
observed by IUE. Their full sample contains 39 quasars, the majority
of which (24) at $z<0.4$, as the quasars discussed in this paper.
Three of the six quasars discussed in this paper are also part of the
Di Clemente et al sample (the three X-ray flat quasars).  Di Clemente
et al. found that the variability amplitude increases with the rest
frame frequency on both 0.3 and 2 years timescales.  Extrapolating
their correlation in the X-ray band one would predict a $<m>$ of 0.75
and 0.83 for the 2 timescales. This is somewhat surprisingly close to
the measured $<m>$ on these timescales, given that variability in the
red band, UV and X-ray may well have a completely different origin.

\section{Discussion}

In our sample of six PG quasars we found clear evidence that X-ray
steep quasars show larger amplitude variations than X-ray flat quasars
on timescales from 2 days to 20 days. On longer timescales we do not
find significant differences between steep and flat quasars, although
the statistics are poorer.  While the sample in this pilot study is
small (and expanded monitoring of a larger sample of quasars is surely
needed), the distinction is so clean cut that we feel justified in
speculating on its origin.

Large narrow band flux variability in sources with a steep spectrum
could be induced by small changes in the spectral index without large
variations of the spectrum normalization. If this is the case each
source should show large spectral variability, with $\alpha_X$
anti-correlated (correlated) with the observed flux if the pivot is at
energies lower (higher) than the observed band. Spectral variability
of this kind has not been observed in the PSPC observations of these
and other quasars (Laor et al. 1997, Fiore et al. 1994) and in NLSy1
(Boller et al 1997, Fiore et al. 1998a, Brandt et al. 1995).  Although
our HRI observations cannot directly rule out this possibility, we
therefore conclude that the observed temporal variability pattern is
not likely the result of complex spectral variability.

Laor et al (1997) suggest that a possible explanation for the
remarkably strong $\alpha_X$--$H_{\beta}$ FWHM correlation is a
dependence of $\alpha_X$ on $L/L_{\rm Edd}$. The line width is
inversely proportional to $\sqrt {L/L_{\rm Edd}}$ if the broad line
region is virialized and if its size is determined by the central
source luminosity (see Laor et al. 1997, \S 4.7).  So narrow-line,
steep (0.1-2 keV) spectrum AGNs emit close to the Eddington luminosity
and have a relatively low mass black hole.  A similar dependence
of spectral shape on $L/L_{\rm Edd}$ is seen in Galactic black
hole candidates (BHC) as they change from the `soft-high' state
to the `hard-low' state.

A physical interpretation for this effect, as described by Pounds et
al. (1995), is that the hard X-ray power-law is produced by
Comptonization in a hot corona and that as the object becomes more
luminous in the optical-UV, Compton cooling of the corona increases,
the corona becomes colder, thus producing a steeper X-ray power-law.
In BHC in `soft-high' states this power law component emerges 
above $\sim 10$ keV, while the spectrum below this energy is 
dominated by softer emission often associated with optically thick 
emission from an accretion
disk. In this model quasar disc emission is at a too low an energy
to observe, since the disc temperature scales with the mass
of the compact object as $M_{BH}^{-1/4}$, leaving the Comptonized
power law to dominate the 2-10 keV spectrum. 
The 0.2-2 keV quasar emission could be due to Comptonization 
(see e.g. Czerny \& Elvis 1987 and Fiore et al. 1995) by a second 
cooler gas component.
The 0.2-2 keV and 2-10 keV spectral indices might well be
correlated with each other, if the emitting regions 
are connected or if the emission mechanisms know about
each other. We have undertaken a campaign of 
observations of the quasars in the Laor et al. (1997) sample
with ASCA and BeppoSAX to clarify this point. Preliminary results
(Fiore et al. 1998b) shows that this is indeed the case: 
steep $\alpha_X$(PSPC) quasars tends to have a steeper hard (2-10 keV) 
X-ray power-law (although the spread of the 2-10 keV
indices seems smaller than that of the PSPC 
indices (as also found by Brandt et al. 1997).

In a sample of quasars with similar luminosities, those
emitting closer to the Eddington luminosity will also be those with
the smaller black hole and hence smaller X-ray emission region.  Thus
light travel time effects would smear intrinsic X-ray variability up
to shorter time scales in high $L/L_{Edd}$ objects, compared with low
$L/L_{Edd}$ objects.  Based on this interpretation, and on
figures \ref{stru_med} and \ref{stru_mean}
it appears that the emission region of the steep soft X-ray
quasars is a factor of $\approx 10$ smaller than that of flat soft
X-ray quasars (about $\approx 10^{16}$ cm and $10^{17}$ cm
respectively).  
Alternatively, the higher variability of steep spectrum
objects may be a true intrinsic property, which could be induced by 
some increased instability in high $L/L_{Edd}$ objects. The range of
luminosity in our sample is too small to tell if the variability amplitude
appears to be better correlated with black hole mass, as expected for
light travel time smearing, or with $L/L_{Edd}$, which would indicate an
intrinsic mechanism.

It is interesting to note that a completely different
analysis, based on the interpretation of the optical to X-ray spectral
energy distribution in terms of emission from accretion disks also
suggest a small black hole mass and an high arretion rate in two NLSy1
(Siemiginowska et al. 1998).

It is also interesting to note that Cyg X-1 in the ``high and soft'' state
(Cui et al. 1997) shows a total root mean square variability higher
than that measured during periods of transitions and in the ``low and
hard'' state.  This is due to strong 1/f noise, extending down to at
least a few $10^{-3}$ Hz, when the source is the ``high and soft''
state (Cui et al. 1997).  When the source is in the ``low and hard''
state this 1/f noise is not present (see e.g. the review of van der
Klis 1995).

Ebisawa (1991) found in a systematic study of Ginga observations of 6 BHC
that the time scales of variability for the soft and hard
components are often different.  The soft component is usually roughly
stable on time scales of 1 day or less, while the hard component
exhibits large variations down to msec time scales.  If the time
scales of BHC and quasars scale with the mass of the compact object
the above two time scales translate to $10^4$ years and 0.1 day
respectively for our quasar sample. For a sample of quasars with
similar redshifts and luminosities this predicts a rather small
scatter in the soft component and a bigger scatter in the hard
component.  ROSAT results go in this direction.  Laor et al. (1994,
1997) find that (for their sample of 23 low--z PG quasars) the scatter
in the normalized 2 keV luminosity is significantly larger than that in the
0.3 keV luminosity.

We conclude that the analogy between AGN and Galactic BHC seems to
hold qualitatively for their X-ray variability properties.

An alternative and intriguing possibility to explain the correlation
between X-ray variability amplitude and spectral shape is that a
component generated closer to the black hole dominates the emission of
steep $\alpha_X$ quasars, as in the spherically converging optically
thick flow proposed by Chakrabarti and Titarchuk (1995) to
characterize BHC in the high and soft state.  If this is the case then
we would again expect that the spectrum of the steep $\alpha_X$
quasars remain steep above 2 keV (and up to $m_ec^2$ according to
Chakrabarti and Titarchuk). Observations with the ASCA and BeppoSAX
satellites instruments, which are sensitive up to 10 keV, (Brandt et
al 1997, Fiore et al 1998b, Comastri et al 1998) suggest that this is
indeed the case.

Schwartz and Tucker (1988) suggested that AGN emission can be produced
by an ensemble of acceleration sites (i.e. shock waves) with different
electron spectral indices and therefore emitting power laws with
different photon indices. In this picture the spectrum is a quadratic
function in log E and the mean index at a given energy arises from the
greatest number of individual acceleration sites. Variability would be
greatest for both steepest energy index and flattest energy index
sources, each of which are dominated by fewer individual regions.  This
is contradicted by the present observations, unless the Soft X-ray
flat quasars become still flatter (and more variable) above 2
keV. High energy X-ray spectroscopy and variability studies are again
needed to obtain a definitive answer. A RossiXTE monitoring campaign
of 4 PG quasars, with a sampling similar to that used in the HRI
campaign, is in progress and could allow us to clarify this point.

\section {Summary}

We have studied the variability of 6 low redshift, radio quiet `PG'
quasars on timescales from a few thousand seconds to years using the
ROSAT HRI and archive ROSAT PSPC observations.  The quasars were
chosen to lie at the two extreme ends of the ROSAT PSPC spectral index
distribution and hence of the H$\beta$ FWHM distribution.  The
observation strategy of the ROSAT HRI compaign has been carefully
designed to provide even sampling on these three basic timescales and
to provide a uniform sampling among the quasars

We have found clear evidence that the X-ray steep, narrow H$\beta$,
quasars show larger amplitude variations than the X-ray flat broad
H$\beta$ quasars on timescales from 2 days to 20 days. On longer
timescales we do not find significant differences between steep and
flat quasars, although the statistics are poorer.  We suggest that the
above correlation between variability properties and spectral
steepness can be explained in a scenario in which the X-ray steep,
narrow optical line objects are in a higher $L/L_{Edd}$ state with
respect to the X-ray flat, broad optical line objects.

Variability on short timescale (a few hundred seconds) is also evident
in the steep $\alpha_X$ quasars. The HRI sensitivity allows to study
the variability of the brightest of these objects down to a few
hundred seconds. To investigate shorter timescale variability 
we will have to await missions with high throughput like AXAF and XMM.

The magnitude of the variability of steep $\alpha_x$ quasar does
not increase on timescales longer than $\approx 10$ days. In particular
for PG1440+356 the power spectrum roughly follows a 1/f law between
$10^{-3}$ and $10^{-6}$ Hz, and below this frequency flattens
significantly indicating a break in the power
spectrum. The frequency of the break is however ill defined,
because of the insufficient sampling on timescales between 1 week and
several months. 

The fact that flat $\alpha_x$ quasars have comparable
variability to steep $\alpha_x$ quasars at $f\sim 10^{-7}$ Hz, but 
significantly smaller variability on $f> 10^{-5}$ Hz suggests that
the variability power spectrum of flat $\alpha_x$ quasars may be similar
in shape to that of steep $\alpha_x$ quasars, but with a break point
scaled to lower frequencies. The other option, that the power spectrum
in flat $\alpha_x$ quasars is just scaled down in amplitude for all $f$, 
is less favoured by our results.
 
A systematic study of a significantly larger sample of quasars
is called for. Such a study would allow one to test the strength
and significance of the $\sigma-\alpha_x$ correlation, to establish
if the differences in variability power spectra is due to a scaling
of the break frequency, or of the overall variability amplitude, 
and to eventually understand if the $\sigma-alpha_x$ correlation is 
driven primarily by $L/L_{\rm Edd}$ or by the black hole mass. The ideal 
instrument for
such a study would be the ROSAT HRI, which is the only X-ray telescope 
which allows the required large number of short exposures.

\bigskip
F.F. acknowledges support from grant NAG5-3039 (ROSAT). M.E. and F.N. 
acknowledge support from ADP grant NAG5-3066.
A.L. acknowledges support by the fund for the promotion of
research at the Technion. We thank Lev Titarchuk and Dan 
Schwartz for useful discussions and comments.



\begin{references}

Boller T., Brandt W.N., Fink H. 1996, A\&A, 305, 53

Boller T., Brandt W.N., Fabian A.C., Fink H. 1997, MNRAS, 289, 393

Brandt W.N., Pounds K.A., Fink H., 1995, MNRAS, 273, L47

Brandt W.N., Mathur S., Elvis M. 1997, MNRAS 285L25

Chakrabarti S.K., Titarchuk L. 1995, ApJ, 455, 623

Comastri A. et al. 1998, A\&A, in press

Cui W., Heindl W.A., Rothshild R.E., Zhang S.N. 
Jahoda K., Focke W. 1997, ApJL, 474, L57

Czerny B., Elvis M. 1987, ApJ, 321, 305

David L.P., Harnden F.R., Kearns K.E., Zombeck M.V. 1997
``The ROSAT High Resolution Imager'', Technical Rep. US ROSAT
Science Data Center/SAO \\
(http://hea-www.harvard.edu/rosat/rsdc\_www/hricalrep.html)

Di Clemente A., Giallongo E., Natali G., Trevese D., Vagnetti F.
1996, ApJ, 463, 466

Ebisawa K., 1991, PhD thesis, ISAS Research Note 483

Fiore F., Elvis M., McDowell J.C., Siemiginowska A., Wilkes B.J.
1994, ApJ, 431, 515

Fiore F., Elvis M., Siemiginowska A., Wilkes B.J., McDowell 
J.C.,  Mathur S., 1995, ApJ, 449, 74

Fiore F., et al. 1998a, MNRAS, in press 

Fiore F., Mineo T., Giallongo E., 1998b, proc. of the Symposium
``The Active Sky'', ed. H. Brandt, F. Fiore, P. Giommi, L. Scarsi,
Rome 21-23 Oct 1997 

Fruscione A., Cagnoni I., Papadakis 1998, ApJ, in preparation

Forster K., Halpern J.P. 1996, ApJ, 468, 565

Kendall M., Stuart A. 1979, The Advanced Theory of Statistics, MacMillan, 
New York

K\"onig M., Staubert R., Timmer J. 1997, proc. of the
conference ``Astronomical Time Series'', Tel Aviv 1-3 Jan 1997

George I., Nandra K., Laor A., Turner T.J., Fiore F., Mushotzky R.F.,
Netzer H. 1997, ApJ, 491, 508

Green A.R., McHardy I.M., Letho H.J. 1993, MNRAS, 265, 664

Groupe D., Beuermann K., Mannheim K., Thomas H.-C.
Fink H.H., de Martino D. 1995, A\&A, 300, L21

Laor A., Fiore F., Elvis M., Wilkes B.J., McDowell J.C. 1994,
ApJ, 435, 611

Laor A., Fiore F., Elvis M., Wilkes B.J., McDowell J.C. 1997,
ApJ, 477, 93

Lawrence A. et al. 1997, MNRAS, 285, 879

Marshall H.L., Carone T.T., Shull J.M.
Malkan M.A., Elvis M. 1996, ApJ 457 169

Mushotzky R., Done C., Pounds K. 1993, Ann. Rev. A\&A, 31, 717

Nandra  K., George I.M., Mushotzky R.F., Turner T.J., Yaqoob T.,
1997, ApJ, 476, 70

Pounds K.A., Done C., Osborne J.P., 1995, MNRAS, 277, L5


Siemiginowska A., Czerny B., 1989, MNRAS 239, 289

Siemiginowska A., Fiore F., Comastri A. et al. 1998, ApJ, in preparation

Schwartz D.A., Tucker W.H., 1988, ApJ, 332, 157

van der Klis, M. 1995, in ``X-ray Binaries'', ed. W.H.G. Lewin,
J. van Paradijs, E.P.J. van den Heuvel, Cambridge Un. press, p. 252 

Zamorani G., et al. 1984, ApJ, 278, 28


\end{references}
\end{document}